\documentclass[preprint,proceedings]{rmaa}

\usepackage{paralist}

\usepackage{psfrag,color}

\newcommand{\Zs}{~Z$_{\odot}$}

\newcommand{\Hb}{\ifmmode {\rm H}\beta \else H$\beta$\fi}
\newcommand{\hii}{H~{\sc ii}}

\newcommand{\Hei}{He~{\sc i} $\lambda$5876}

\newcommand{\Heii}{He~{\sc ii} $\lambda$4686}
\newcommand{\Nii}{[N~{\sc ii}] $\lambda$6584}

\newcommand{\Oi}{[O~{\sc i}] $\lambda$6300}

\newcommand{\Oii}{[O~{\sc ii}] $\lambda$3727}

\newcommand{\Oiii}{[O~{\sc iii}] $\lambda$5007}
\newcommand{\Oiiit}{[O~{\sc iii}] $\lambda$4363}

\SetYear{2002}
\SetConfTitle{TexMex 8}

\title{About the emission line sequence of H II galaxies} 

\author{
  Gra\.{z}yna  Stasi\'{n}ska\altaffilmark{1}, 
  Yuri Izotov\altaffilmark{2}}

\altaffiltext{1}{LUTH, Observatoire de Meudon, Meudon, France.}
\altaffiltext{2}{Main Astronomical Observatory,
Kyiv,  Ukraine.}

\shortauthor{Stasi\'{n}ska \& Izotov}
\shorttitle{The emission line sequence of H {\sc ii}
galaxies}

\fulladdresses{
\item  Gra\.{z}yna  Stasi\'{n}ska: 
LUTH, Observatoire de Meudon, 5 Place Jules Janssen,
                      F-92195 Meudon Cedex, France 
                      (grazyna.stasinska@obspm.fr).
\item Yuri Izotov: Main Astronomical Observatory,
Ukrainian National Academy of Sciences,
Kyiv 03680,  Ukraine (izotov@mao.kiev.ua).}

\listofauthors{ G.  Stasi\'{n}ska \& Y. Izotov}
\indexauthor{Stasi\'{n}ska, G.}
\indexauthor{Izotov,Y.}

\abstract{ 

We consider 400 \hii\ galaxies in which the oxygen abundances 
were obtained by electron temperature based methods. We split the sample 
in three metallicity bins. In each bin, 
 a narrow sequence is found not only 
in pure emission line ratio diagrams 
but also in terms of \Hb\ equivalent 
width, indicating the existence of an evolutionary sequence. Our 
diagrams show unambiguously the existence of an evolution on 
a timescale of a few Myr.
We compare the observed sequences with  
photoionization models of
appropriate metallicity. In order to understand the 
evolution of \hii\ galaxies one needs to consider the evolution of the 
gas as well as that of the stars. The observations require EW(\Hb) 
to decrease more rapidly than 
predicted by the passive evolution of a starburst. A
photoionized adiabatic expanding bubble with 
a covering factor decreasing with time reproduces 
most diagrams. However, the question of the heating of \hii\ 
galaxies and the origin of the nebular \Heii\ emission are not 
settled.
We find evidence for self-enrichment in nitrogen on a time scale of 
several Myr.  }

\addkeyword{H~II regions}

\begin{document}
\maketitle

\section{Introduction}

It has been known for years that giant \hii\ regions form a narrow sequence 
in various emission line ratio diagrams (e.g. McCall,
Rybski \& Shields 1985). The reasons for the existence of 
such a sequence are not yet fully understood. 
 The first interpretations, based on single star photoionization 
 models,concluded that 
metallicity is driving the sequence,
 but variation of an additional parameter -- either
the effective temperature
of the stars or the ionization parameter -- is needed 
to reproduce the observed sequence (McCall et al. 1985; Dopita
\& Evans 1986). More recent studies using stellar population synthesis 
have shown that the spectral 
energy distribution of the ionizing radiation
depends naturally on metallicity, both 
 through stellar structure and evolution and through
the effect of opacities in the 
stellar atmospheres (e.g. Mc Gaugh 1991). 

However,  star formation proceeds by
bursts and giant \hii\  regions are powered by clusters of
coeval stars (e.g. Sargent \& Searle 1970, Mas-Hesse \& Kunth 1999). 
Stellar evolution produces a gradual modification of the
integrated stellar energy distribution.  Spectra of \hii\ regions
whose ionization is dominated by one single cluster can then be caught
at different cluster ages.  This is the case 
of blue compact dwarf galaxies or dwarf irregular galaxies, which are 
referred to under the common name of \hii\ galaxies.  These objects 
make up invaluable tools to study the evolution of
stellar systems and their interaction with the surrounding
interstellar medium during the phase where massive O stars are
present.

\begin{figure*}[!t]\centering

 \includegraphics[width=0.7\textwidth]{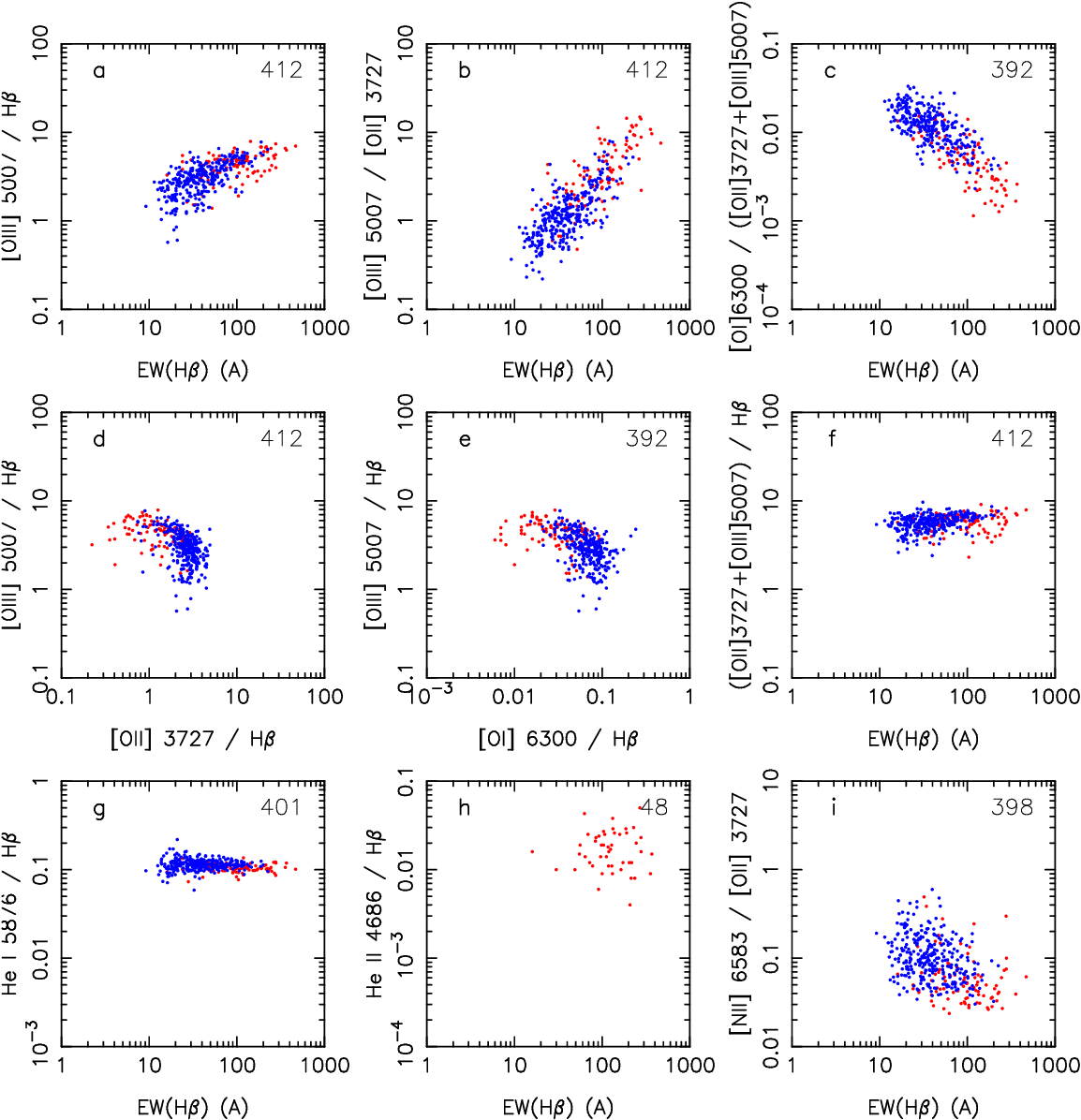}
\caption{The entire sample of \hii\ galaxies in 
diagrams involving line ratios and $EW$(\Hb). The number of data
points is shown in  each 
panel. Byurakan objects are in red, SDSS objects are 
in blue.
}
  \label{fig1}
\end{figure*}

Several studies (see references in Stasi\'{n}ska \& Izotov
2003, hereafter SI2003) have been devoted to this question, based on various
samples and somewhat different approaches.  Here, we give an
account of the SI2003 paper, which aimed at finding what
conditions are needed to reproduce the observed sequence of \hii\
galaxies with models of \hii\ regions surrounding evolving starbursts. 
The success of this enterprise has greatly benefited from the fact
that we were able to construct a homogeneous sample of \hii\ galaxies
of unprecedented size, in which the oxygen abundances were derived in a
model-independent way.  This allowed us to divide the sample into
three metallicity bins, and study each of them independently.

We describe the data sample, comment on the 
observational trends and present some model sequences.
 Details about the sample and the modeling 
procedure can be found in SI2003. We take the opportunity of this 
conference to develop on some aspects that 
were only briefly mentioned in SI2003. In particular, 
we refer to the viewpoint presented by R. Terlevich at this conference.

\section{The observational sample}

Ideally, the sample should be as much as possible relevant, 
homogeneous,  complete (or with well understood biases) and
large (since there are at least two independent parameters:
age and metallicity).

Metal-poor \hii\ galaxies are the best candidates to 
study the evolution of giant \hii\ regions. As opposed to giant 
\hii\ regions in spiral galaxies, their spectra are not affected by 
the light from the stars located in the disk. Also,
the high enough electron temperature allows detection 
of the \Oiiit\ line and therefore direct 
determination of the oxygen abundance by the electron 
temperature method (this is not the case of giant \hii\ regions found 
in the inner parts of disk galaxies). In addition, 
the dust content in \hii\ galaxies is small, which makes the 
interpretation of the data easier and more robust. 

\begin{figure*} [!t]\centering
     \includegraphics[width=0.7\textwidth]{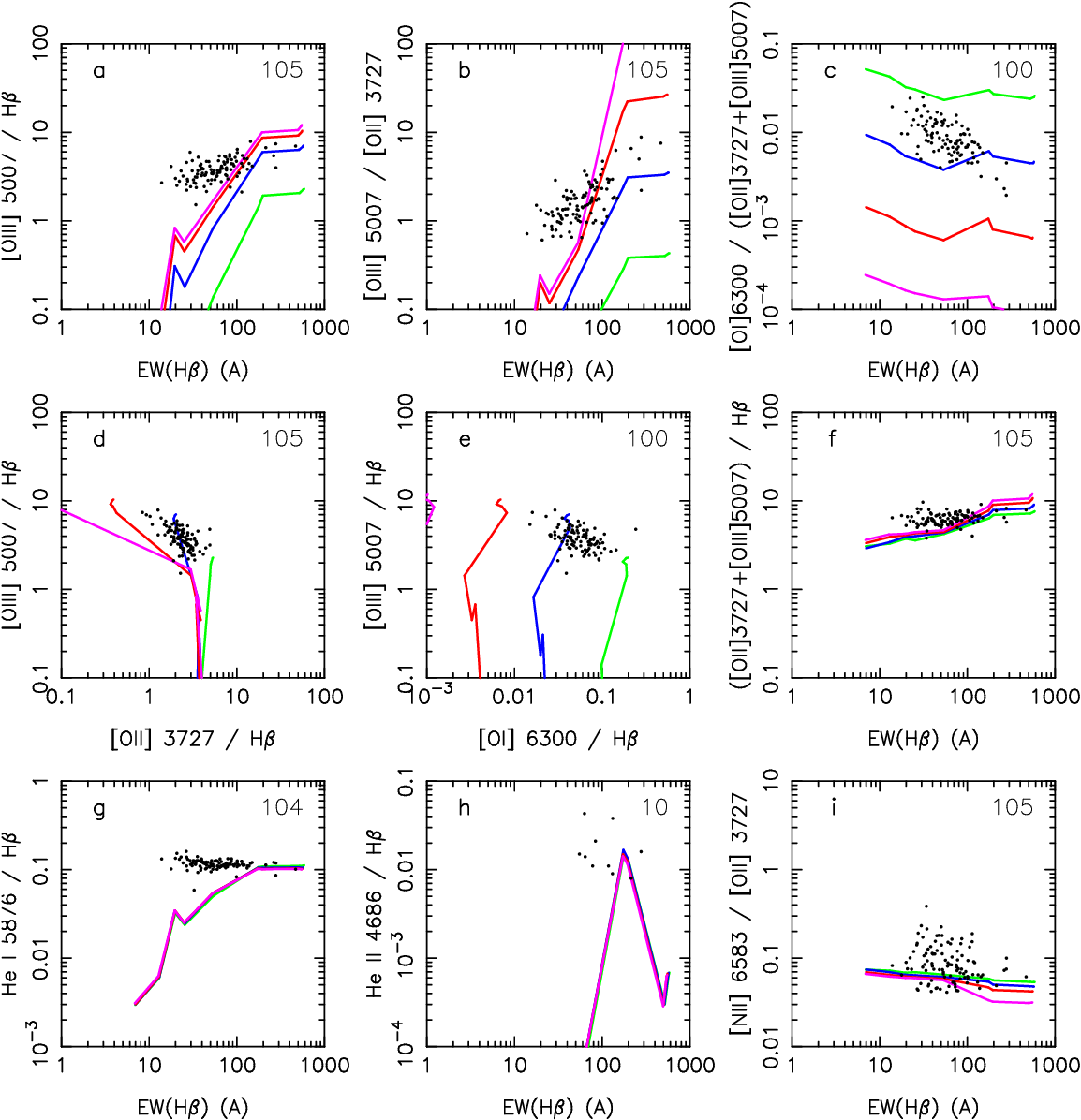}
  \caption{The ``high'' metallicity bin. The data points correspond to
the objects of our sample for which the estimated error in the \Oiiit\
intensity is less than 50\%  and which have  O/H $>
10^{-4}$. The number of data
points present in each diagram is shown in the upper right of each panel.
Overplotted are evolutionary sequences of models with
metallicity  $Z$ = 0.2\Zs and constant density. Each sequence 
corresponds to a different initial ionization parameter. 
 }
  \label{fig2}
\end{figure*}

Our sample of \hii\ galaxies was obtained by merging 
   $\sim$ 100 blue compact
galaxies from the First and Second Byurakan
surveys  and $\sim$ 300 emission-line galaxies from the
early data release of the SLOAN digital sky survey (SDSS,
Stoughton et al. 2002).  Only objects with a
detected [O {\sc iii}] $\lambda$4363 emission line were included. 
The spectra of the entire sample were reduced and reddening-corrected 
with identical  procedures and the abundances were derived in the same 
way.

The spectra of the  Byurakan survey galaxies were obtained 
by Izotov and coworkers
 to determine the pregalactic helium abundance, therefore the
signal-to-noise is high and  
the \Hb\ equivalent widths, $EW$(\Hb), are large by observer's 
selection. The SDSS galaxies reach larger 
distances, but  their  \Hb\ luminosities are rougly similar to those 
of the Byurakan galaxies. For the majority of the objects,
       the extinction is small ($E(B-V) < 0.2$).

\section{Diagnostic diagrams}

\begin{figure*} [!t]\centering
         \includegraphics[width=0.7\textwidth]{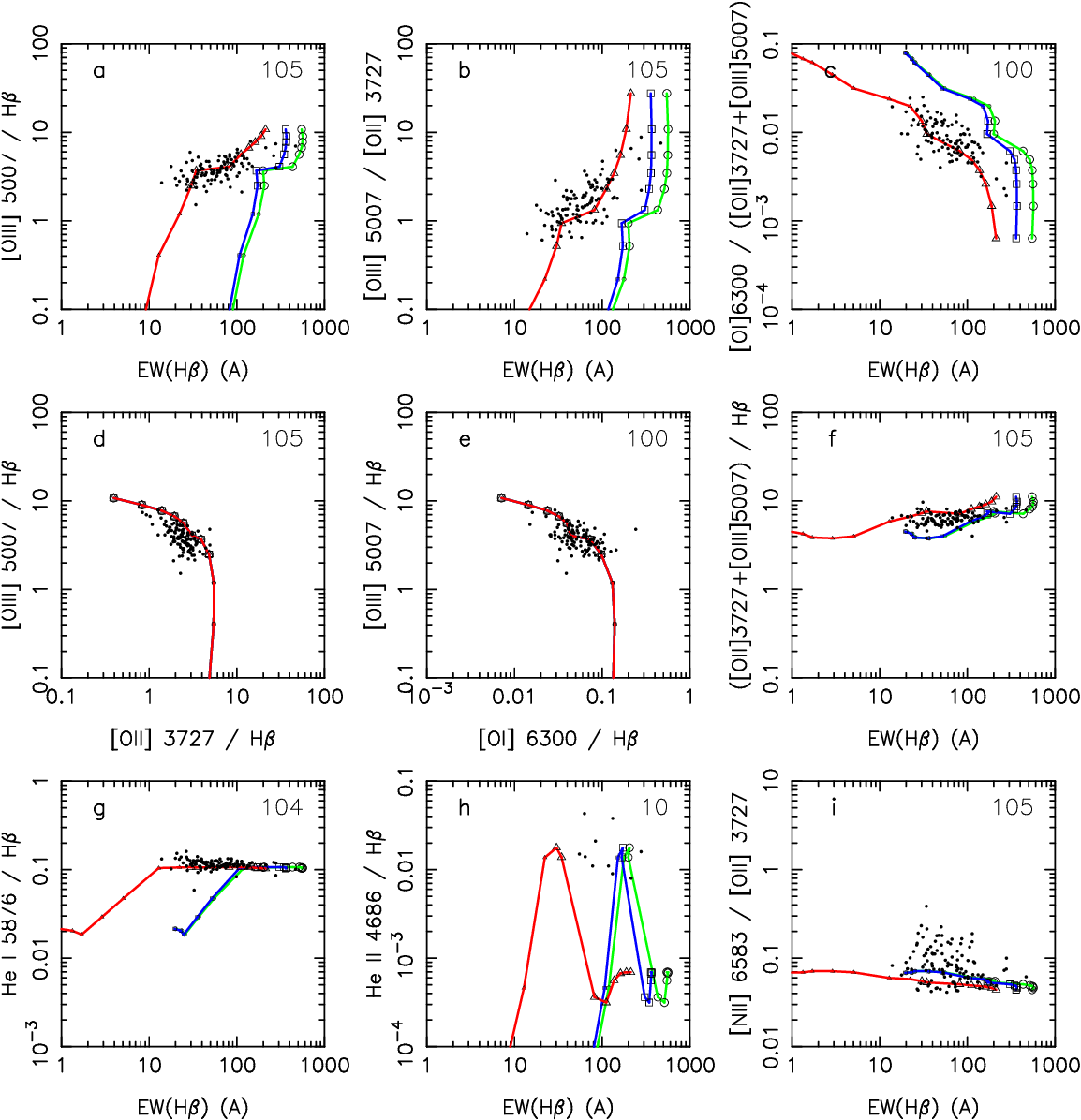}
  \caption{The ``high'' metallicity bin.  Same data points as 
  in Fig. 2.
Overplotted are evolutionary sequences of expanding bubble models with
metallicity $Z$ = 0.2\Zs.  The green curve represents the pure
expanding bubble sequence.  The blue curve is obtained by adding an 
underlying older population.  The red curve is obtained from
the blue one assuming a covering factor decreasing with time.}

  \label{fig3}
\end{figure*}

The  diagrams were carefully chosen to provide strong
constraints on the evolution of the ionizing clusters and of their 
surrounding \hii\ regions. The relatively high quality of the data for 
the entire sample allowed us to use not only 
the classical strong lines, i.e. \Oii, \Oiii, and \Hb, but also 
weaker lines such as \Oi, \Hei\ and \Heii\ (the latter, being at the level 
of 0.003 -- 0.03 of \Hb, is detected only in the Byurakan sample). 
These weak lines should not be discarded as they provide very 
important diagnostics. 
We also consider a diagram involving \Nii/\Oii. 

Fig.  1 shows the distribution of the observational points in 9
diagrams.  The relative distribution of the Byurakan 
objects (red points) and of the SDSS objects (blue points) 
implies that the nature of the objects is likely similar
and justifies the merging of the two samples.
Fig. 1 shows that \hii\ galaxies form a sequence not only in the 
classical line-ratio diagrams (\Oiii/\Hb\ vs \Oii/\Hb\ and \Oiii/\Hb\ vs 
\Oi/\Hb\ \footnote{the sequence in the pure line-ratio diagrams 
is truncated with respect to diagrams 
built with \hii\ regions from spiral galaxies, since the 
metal-rich objects are lacking.}) but also, and very clearly, 
as a function of  $EW$(\Hb).
This indicates an evolutionary sequence: either on a timescale of a few
 Myr, as the death of massive stars progressively reduces 
 the \Hb\ emission or on
 time scales of Gyr, as previous generations of stars gradually build up
 the \Hb\ continuum. The latter view is the one adopted by Terlevich 
 in these proceedings, using different arguments 
\footnote{The sample used by Terlevich contains also metal-rich objects
 where the effects of previous generations of stars are expected to be
 stronger.}. The fact that the \Oiii/\Oii\ and 
 \Oi/(\Oii+\Oiii) ratios show very strong trends with $EW$(\Hb) 
 and that the same trends are seen for each metallicity bin (see 
 Figs. 2 and 4 and SI2003 for more details) argue for the first 
 interpretation as being  the main cause of the observed sequence 
in our sample. 
 Therefore, $EW$(\Hb) is indicative of the age on the ionizing 
 stellar population (although not directly, since several factors
 must be taken into account, see later).

The fact that a sample of 400 objects shows such clear trends is
remarkable.  This means that, in spite of the complexity revealed by 
detailed studies of a few  \hii\ galaxies, 
the overall evolution of such objects must obey simple laws.

Note that  [N~{\sc
ii}] $\lambda$6584/[O~{\sc ii}] $\lambda$3727 is seen to increase as
$EW$(H$\beta$), although with higer dispersion than  
 \Oi/(\Oii+\Oiii).

\begin{figure*} [!t]\centering
 \includegraphics[width=0.7\textwidth]{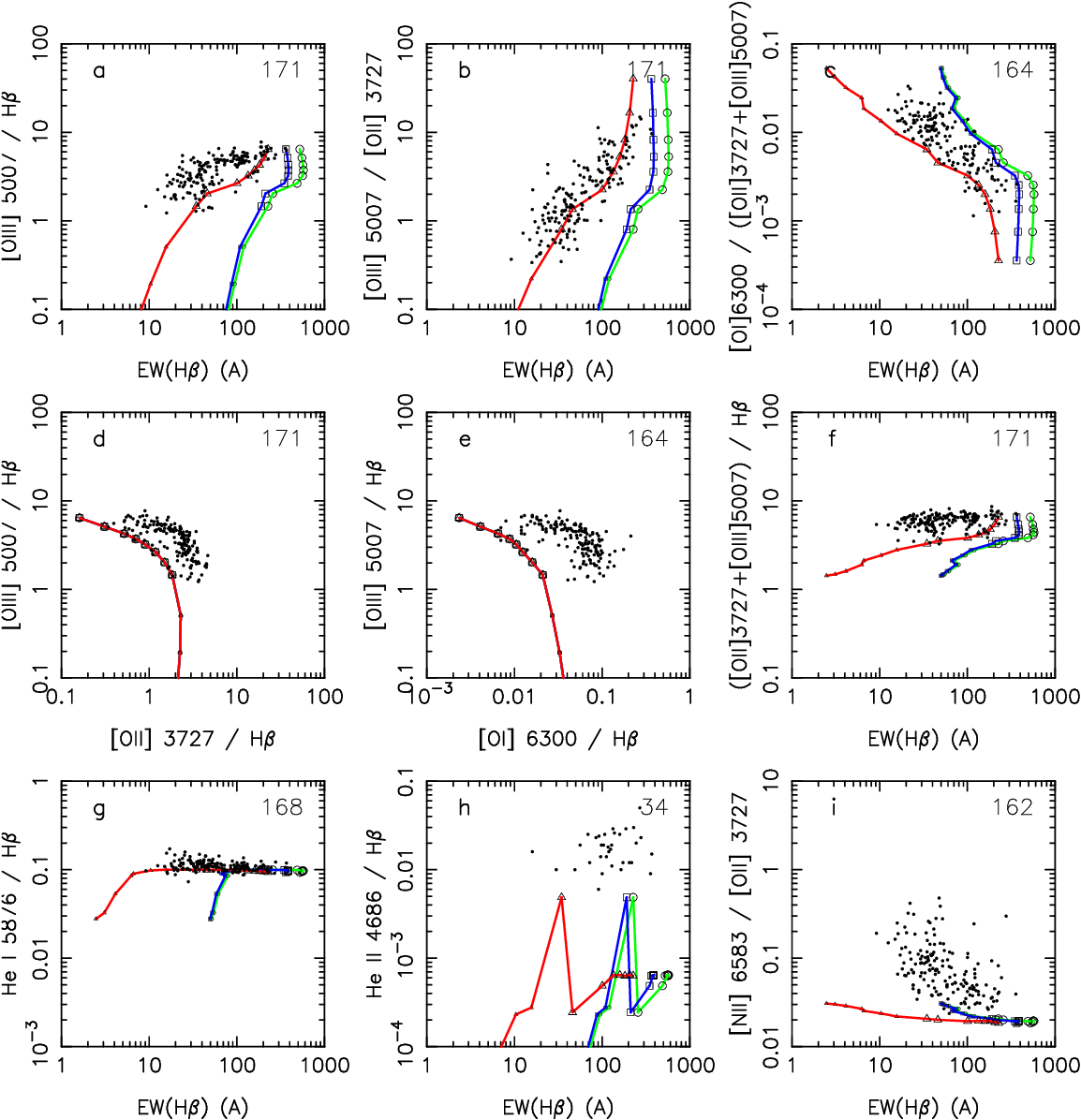}
  \caption{Same as Fig. 3, for the ``intermediate'' metallicity bin.
}
  \label{fig4}
\end{figure*}

\section{The modeling strategy}

All the details are published in SI2003. The observational 
sample is subdivided in 3 metallicity bins: `'high'' metallicity 
 (between 0.25 and  0.12 times solar),``intermediate'' (between 0.12 and
 0.036 times
solar) and `'low'' (below 0.036 times solar). For each bin we 
construct sequences of photoionization models that represent the evolution 
of a giant  \hii\ region of appropriate metallicity. The aim is to find 
what conditions are needed to reproduce simultaneously all the 
observed diagrams (within the known uncertainties and biases) both 
from the point of view of the location of the sequences and of the 
density of points in each zone of the diagrams. We start
with the simplest physically reasonable models and proceeded step by step, 
justifying the increase in complexity by the requirement of fitting 
the data.

\begin{figure*} [!t]\centering
 \includegraphics[width=0.7\textwidth]{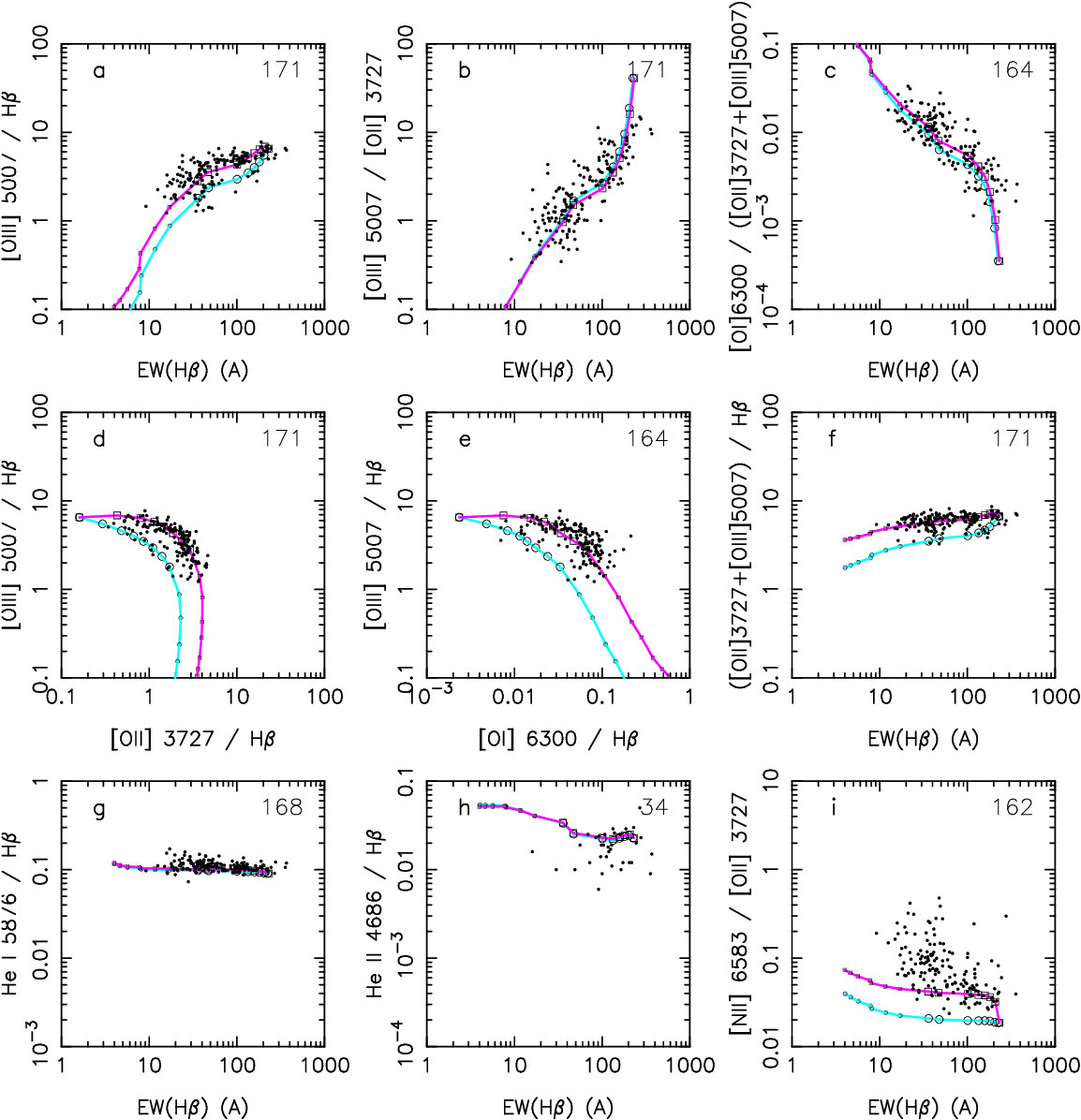}
  \caption{The ``intermediate'' metallicity bin
as in Fig. 4. The cyan curve corresponds to the same sequence as 
the red curve in Fig. 4 but with X-rays added. 
The purple curve corresponds to the same sequence but for a 
two-component chemical composition of the gas.}
  \label{fig5}
\end{figure*}

\section{Results}

 Fig. 2 shows that the 
classical modeling with non-evolving  homogeneous spheres 
 is incapable of reproducing the 
observed slopes in panels a, b and c. But the evolution of the exciting 
stars is not the only factor modifying the spectrum 
of a giant \hii\ region.

One of the most striking observed trends is the steady increase of
  \Oi/(\Oii+\Oiii) as $EW$(H$\beta$) decreases. 
As known, stellar winds sweep the 
surrounding matter and confine the   emitting gas in a thin expanding shell.  
This induces  an important decrease of the ionization 
parameter and allows to reach the highest observed values of  \Oi/(\Oii+\Oiii)
 in about 5~Myr (see green curve in Fig. 3).  However, it does not 
 reproduce the observed 
 slopes in panels a, b, c. Also, the models give higher  $EW$(\Hb) than 
 observed in early ages. 
Selective dust absorption, if existing, would have only 
 a minor effect since the extinction is small. 
 The presence of an older, non ionizing
stellar population,  
attested by  stellar features,
 increases the continuum and thus reduces $EW$(\Hb).  
However, 
 the older population is independent of the age of the most recent 
 starburst (although it may depend on the metallicity, being likely 
 more important for \hii\ galaxies having reached a higher metallicity). 
 The blue curve in Figs. 3 and 4 represents a sequence with 
 an average old population added. While this brings the 
 highest $EW$(\Hb) in better agreement with the observations, this does not 
 solve the problem of the slopes in panels a,  b and c, and does not 
 explain why \Hei/\Hb\ remains constant in the entire range of $EW$(\Hb).
What is needed is to lower  $EW$(\Hb) more rapidly.  One way to achieve this 
is to consider a covering factor of the ionizing source decreasing
with time.

While this can explain the ``high'' metallicity bin,
 the ``intermediate'' and ``low'' metallicity bins 
show further problems.  
The He~{\sc ii} $\lambda$4686 nebular emission 
occurs too frequently and in too wide a range of $EW$(H$\beta$) to be
attributed to either the hard photons from Wolf-Rayet
stars (Schaerer 1998), or the X-rays
from young starbursts
(Cervi\~{no} et
al. 2002). The cyan curve in Fig. 5 shows the effect of photoionization 
by an X-ray source of 
luminosity  $4\times 10^{40}$~erg~s$^{-1}$ 
(10 times the initial \Hb\ luminosity). Such X-rays
could originate from 
massive binaries or from supernova remnants 
 in the age range of 10 -- 50~Myr (Van Bever \& Vanbeveren 2000).
Even this  amount of X-rays cannot heat the gas 
enough
to reconcile models and observations in panels d and e.   
Additional heating mechanisms must be investigated.  Hard photons 
from planetary nebulae and white dwarfs from earlier
generations of stars are certainly not enough. Photoelectric 
heating by dust grains is likely negligible, due to the low ionization 
parameters involved. Shocks produced by winds and supernovae 
from the latest burst of star formation are far from sufficient. But the
hypothesis of either turbulent heating or shock heating due to stellar winds from \emph{previous}
generations of stars  are promising and need to be
investigated. However, a heating deficiency is not the only way to 
explain the discrepancy. 
Points in  line ratio diagrams can also be displaced by the 
presence of a diffuse gas component. 
Another option is to invoke chemical inhomogeneities, which are indeed
 expected in zones affected by mass loss and
supernova explosions from massive stars. The question is what is the state 
of the newly synthesized matter, and how much of it has escaped
the nebula. The problem is far from settled.
Direct evidence for self-enrichment is scarce
(see Kobulnicky 1999) but the observed increase of
\Nii/\Oii\ as $EW$(H$\beta$) decreases argues for
a nitrogen self-enrichment  on timescales of a few
Myr. Of course, this in itself does not argue for any oxygen self-enrichment,
 since the nitrogen and oxygen donors are not the same.
As an example, the purple curve in Fig. 5 combines two sequences with nebular
metallicities 0.2~Z$_{\odot}$ and 0.05~Z$_{\odot}$ (the apparent 
metallicity of these combined models as derived from temperature 
based methods would of course be intermediate between 
these values) and is good agreement with the observations.
 Similar conclusions are drawn for the `'low'' 
metallicity bin.

\section{Final remarks}

Considering a large and homogeneous sample of \hii\ galaxies with 
oxygen abundances determined from temperature-based methods, 
we have built three subsamples of different metallicities and compared 
each of them with models of \hii\ regions ionized by evolving star 
clusters of appropriate metallicity. This allowed us 
to distinguish evolution and abundance effects in the 
emission line sequence of \hii\ galaxies. This, and the use of weak lines 
in conjunction with classical strong lines 
allowed us to sharpen up our view on the evolution of  \hii\ galaxies.
\hii\ galaxies draw a tight
 sequence not only in metallicity but also in age, on a time scale 
of a few Myr, corresponding to the lifetimes of massive stars. This is in 
agreement with the view that the excitation of an \hii\ galaxy is 
dominated by a cluster of coeval massive stars. 
We have shown that unveiling the evolution of \hii\ 
galaxies requires taking into account the dynamical evolution of the gas. 
The adiabatic expanding bubble theory provides a good frame. 
While the equivalent width of \Hb\ is definitely linked to the age of 
the most recent 
starburst,  the observations demand that $EW$(\Hb) decreases faster 
 than predicted by normal photoionization models. This 
cannot be the effect of an older stellar population alone or of
selective dust absorption, and calls for a covering factor decreasing 
with time (implying that a growing amount of Lyman continuum 
radiation is escaping from \hii\ galaxies). 
\adjustfinalcols
The origin of the nebular \Heii\ emission  is not clear. 
The question of the heating of giant \hii\ regions is not 
definitely settled.  We see evidence for an enrichment in nitrogen on a 
timescale of a few Myr. Possibly, \hii\ galaxies are not homogeneous 
in oxygen abundance. 

While our study has allowed to considerably improve our diagnostic on 
the evolution of \hii\ galaxies, a real understanding of this 
evolution now implies a dynamical modeling that would be able to 
reproduce all the observed trends.

\begin{acknowledgements}
G. S. thanks  the conference organizers for support.
Y. I. acknowledges support from
the Observatoire de Paris
and the Swiss SCOPE 7UKPJ62178 grant.
The Sloan Digital Sky Survey is a joint project of The
University of
Chicago, Fermilab, the Institute for Advanced Study, the Japan
Participation
Group, the Johns Hopkins University, the Los Alamos National
Laboratory,  
the
Max-Planck-Institute for Astronomy (MPIA), the Max-Planck-Institute
for
Astrophysics (MPA), New Mexico State University, Princeton
University, the
United States Naval Observatory, and the University of Washington.
Funding for the project has been provided by the Alfred P. Sloan
Foundation,
the Participating Institutions, the National Aeronautics and Space
Administration, the National Science Foundation, the U.S. Department
of Energy,
the Japanese Monbukagakusho, and the Max Planck Society.

\end{acknowledgements}

\end{document}